\documentclass[12pt]{article}
\usepackage{fullpage}
\usepackage{amssymb}
\usepackage{amsmath}
\usepackage{euscript}
\usepackage{graphicx}
\usepackage{times}
\usepackage{color}
\newtheorem{theorem}{Theorem}
\newtheorem{example}{Example}

\newtheorem{lemma}{Lemma}
\newtheorem{corollary}{Corollary}

\newtheorem{definition}{Definition}
\newtheorem{proposition}{Proposition}

\newcommand{\beq}{\begin{equation}}
\newcommand{\eeq}{\end{equation}}
\newcommand{\be}{\begin{enumerate}}
\newcommand{\ee}{\end{enumerate}}
\newcommand{\bi}{\begin{itemize}}
\newcommand{\ei}{\end{itemize}}
\newcommand{\bd}{\begin{description}}
\newcommand{\ed}{\end{description}}
\newcommand{\bc}{\begin{center}}
\newcommand{\ec}{\end{center}}
\newcommand{\bthm}{\begin{theorem}}
\newcommand{\ethm}{\end{theorem}}
\newcommand{\bdefi}{\begin{definition}}
\newcommand{\edefi}{\end{definition}}
\newcommand{\bcor}{\begin{corollary}}
\newcommand{\ecor}{\end{corollary}}
\newcommand{\blem}{\begin{lemma}}
\newcommand{\elem}{\end{lemma}}
\newcommand{\bexa}{\begin{example}}
\newcommand{\eexa}{\end{example}}
\newcommand{\bprop}{\begin{proposition}}
\newcommand{\eprop}{\end{proposition}}

\newcommand{\remove}[1]{}

\def\real{\hbox{\rm\setbox1=\hbox{I}\copy1\kern-.45\wd1 R}}
\def\neal{\hbox{\rm\setbox1=\hbox{I}\copy1\kern-.45\wd1 N}}

\def\argmax{\mbox{\rm \em argmax}}

\begin{document}

\begin{titlepage}
\title{Approximately Optimal Mechanism Design \\ via Differential Privacy%
\thanks{We thank Amos Fiat and Haim Kaplan for discussions at an early stage of this research. We thank Frank McSherry and Kunal Talwar for helping to clarify
issues related to the constructions in~\cite{MT07}. Finally, we thank Jason
Hartline, James Schummer, Roberto Serrano and Asher Wolinsky for their valuable comments.}}

\author{Kobbi Nissim\thanks{Microsoft Audience Intelligence, Israel,
and Department of Computer Science, Ben-Gurion University. Research
partly supported by the Israel Science Foundation (grant No.\
860/06). {\tt kobbi@cs.bgu.ac.il}.}
\and Rann
Smorodinsky\thanks{Faculty of Industrial Engineering and Management,
Technion -- Israel Institute of Technology, Haifa 32000, Israel. This
work was supported by Technion VPR grants and the Bernard M. Gordon
Center for Systems Engineering at the Technion. {\tt
rann@ie.technion.ac.il}.} \and Moshe Tennenholtz\thanks{Microsoft
Israel R\&D Center and the Faculty of Industrial Engineering and
Management, Technion -- Israel Institute of Technology, Haifa 32000,
Israel. {\tt moshet@microsoft.com}.}}
\date{}
\maketitle

\begin{abstract}
In this paper we study the implementation challenge in an abstract interdependent
values model and an arbitrary objective function. We design a mechanism that allows for approximate optimal implementation of {\em insensitive} objective functions in ex-post Nash equilibrium. If, furthermore, values are private then the same mechanism is strategy proof. We cast our results onto two specific models: pricing and facility location. The
mechanism we design is optimal up to an additive factor of the order
of magnitude of one over the square root of the number of agents 
and involves no utility transfers.

Underlying our mechanism is a lottery between two auxiliary
mechanisms --- with high probability we actuate a mechanism that reduces players influence on the choice of the social alternative, while choosing the optimal outcome with high probability.
This is where the recent notion of {\em differential privacy} is employed. With the complementary probability we actuate a mechanism that is typically far from optimal 
but is incentive compatible.
The joint mechanism inherits the desired properties from both.

\end{abstract}

\thispagestyle{empty}

\end{titlepage}

\renewcommand{\thefootnote}{\arabic{footnote}}
\setcounter{footnote}{0}

\section{Introduction}

Mechanism design deals with the implementation of desired outcomes
in a multi-agent system with asymmetric information. The outcome of
a mechanism may be a price for a good, an allocation of goods to the
agents, the decision on a provision of a public good, locating
public facilities, etc. The quality of the outcome is measured by
some objective function. In many instances the literature is
concerned with the sum of the agents' valuations for an outcome, but
the objective function can take many other forms, such as the
revenue of a seller in an auction setting, the social inequality in
a market setting and more. The reader is referred to Mas-Colell,
Whinston and Green ~\cite{mascolell} for a broader introduction. The
holy grail of the mechanism design challenge is to design mechanisms
which exhibit dominant strategies for the players, and furthermore,
once players play their dominant strategies the outcome of the
mechanism coincides with maximizing the objective function. Broadly
speaking, this challenge is equivalent to designing optimal direct mechanisms
that are truthful.

As it turns out, such powerful mechanisms do not exist in general.
The famous Gibbard-Satterthwaite theorem (Gibbard~\cite{Gibbard} and
Satterthwaite~\cite{Satterthwaite}) tells us that for non-restricted
settings any non-trivial truthful mechanism is dictatorial. However,
if we restrict attention to the objective function that is simply
the sum of the agents' valuations, then this problem can be overcome
by introducing monetary payments. Indeed, in such cases the
celebrated Vickrey-Clarke-Groves mechanisms, discovered by
Vickrey~\cite{Vickrey} and generalized by Clarke~\cite{Clarke} and
Groves~\cite{Groves}, guarantee that being truthful is a dominant
strategy and the outcome is optimal. Unfortunately,
Roberts~\cite{Roberts} showed that a similar mechanism cannot be
obtained for other objective functions. This cul-de-sac induced
researchers to `lower the bar' for mechanism design. One possibility
for lowering the bar is to replace the solution concept with a
weaker one and a large body of literature on Bayes-Nash
implementation has developed (the reader is referred to Mas-Colell
et al. \cite{mascolell} for further reading).

Another direction is that of approximate implementation where the quest replaces accurate
implementation with approximate implementation, while keeping the
approximation inaccuracy as low as possible. The latter research agenda turned out to be fruitful and yielded
many positive results. A sequence of papers on {\em virtual
implementation}, initiated by Matsushima \cite{Matsushima88} and
Abreu and Sen \cite{Abreu and Sen 90}, provides general conditions
for approximate implementation where the approximation inaccuracy in
a fixed model can be made arbitrarily small. On the other hand, the
recent literature emerging from the {\em algorithmic} mechanism
design community looks at approximation inaccuracies which are a
function of the {\em size} of the model (measured, e.g., the number of
agents).

Interestingly, no \emph{general} techniques are known for designing
mechanisms that are approximately optimal for arbitrary social
welfare functions. To demonstrate this consider the facility
location problem, where a social planner needs to locate some
facilities, based on agents' reports of their own location. This
problem has received extensive attention recently, yet small changes
in the model result in different techniques which seem tightly
tailored to the specific model assumptions (see Alon et
al.~\cite{DBLP:journals/corr/abs-0907-2049}, Procaccia and
Tennenholtz~\cite{DBLP:conf/sigecom/ProcacciaT09} and Wang et
al.~\cite{facility10}).

Another line of research, initiated by Moulin~\cite{Moul80}, is that on mechanism design {\em without money}. Moulin, and later Schummer and
Vohra~\cite{SV04,SV07}, characterized
functions that are truthfully implementable without payments and
studied domains in which non-dictatorial functions can be
implemented. More recently, Procaccia and Tennenholtz~\cite{DBLP:conf/sigecom/ProcacciaT09}
studied a relaxation of this notion -- {\em approximate} mechanism design without money.

Our work presents a general methodology for designing approximately optimal mechanisms for a broad range of models, including the facility location problem. A feature of our constructions is that the resulting mechanisms do not involve monetary transfers.

\subsection{Our Contribution}

We introduce an abstract mechanism design model where agents have
interdependent values and provide a generic technique for
approximate implementation of an arbitrary objective function. More
precisely, we bound the worst case difference between the optimal
outcome (`first best') and the expected outcome of our generic
mechanism by $O(\sqrt{\frac{\ln n}{n}})$, where $n$ is the
population size. In addition, our generic construction does not
involve utility transfer.

Our construction combines two very different random mechanisms:
\begin{itemize}
\item With high probability we deploy a mechanism that chooses social alternatives
with a probability that is proportional to (the exponent of) the
outcome of the objective function,
assuming players are truthful.
This mechanism exhibits two important properties. First, agents
have small influence on the outcome of the mechanism and
consequently have little influence on their own utility. As a result
all strategies, including truthfulness, are $\epsilon$-dominant.
Second, under the assumption that players are truthful, alternatives
which are nearly optimal are most likely to be chosen. The concrete
construction we use follows the Exponential Mechanism presented by
McSherry and Talwar~\cite{MT07}.

\item With vanishing probability we deploy a mechanism which is designed with the goal of
eliciting agents' private information, while ignoring the objective function.

\end{itemize}

Our technique is developed for settings where the
agents' type spaces as well as the set of social alternatives are
finite. In more concrete settings, however, our techniques extend to
`large' type sets. We demonstrate our results in two specific settings: (1) Facility location problems,
where the social planner is tasked with the optimal location of $K$
facilities in the most efficient way. In this setting we focus on
minimizing the social cost which is the sum of agents' distances
from the nearest facility. (2) The digital goods pricing model,
where a monopolist needs to determine the price for a digital good
(goods with zero marginal cost for production) in order to maximize
revenue.

Another contribution of our work is an extension of the classical
social choice model. In the classical model agents' utilities are
expressed as a function of the private  information and a social
alternative, a modeling that abstracts away the issue of how agents
exploit the social choice made. We explicitly model this by extending the standard model by an additional stage, following
the choice of the social alternative, where agents take an action to exploit the social alternative and determine their utility (hereinafter `reaction'). We motivate
this extension to the standard model with the following examples:
(1) In a Facility Location problem agents react to the mechanism's
outcome by choosing one of the facilities (e.g., choose which school to attend). (2) A Monopolist posts a
price based on agents input. Agents react by either buying the good
or not. (3) In an exchange economy agents react to the price vector
(viewed as the outcome of the invisible hand mechanism) by demanding
specific bundles. (4) In a public good problem, where a set of
substitutable goods is supplied, each agent must choose her favorite
good. (5) Finally, consider a network design problem, where each
agent must choose the path it will use along the network created by the society. These
examples demonstrate the prevalence of
`reactions' in a typical design problem.%
\footnote{Formally, the introduction of reactions only generalizes
the model. In fact, if we assume that the set of reactions is a
singleton then we are back to the classical model. Additionally, it
could be argued that reactions can be modeled as part of the set
social alternatives, $S$. For the analysis and mechanism we propose
the distinction between the set $S$ and the reactions is important.}
With this addendum to the model one can enrich the notion of a
mechanism; in addition to determining a social choice the mechanism
can also restrict the set of reactions available to an agent. For
example, in the context of school location, the central planner can
choose where to build new schools and, in addition, impose the
specific school assigned to each student. We refer to this aspect of
mechanisms as {\it imposition}.

We demonstrate the notion of imposition with the following illustrative example:

\bexa In time of depression the government proposes to subsidize
some retraining programs. There are three possible programs from
which the government must choose two due to budget constraints. Once
a pair of programs is chosen each agent is allocated to her favorite program.
For simplicity, assume each candidate for retraining has a
strict preference over the three programs, with utilities equal
$1,2$ and $3$. Assume the government wants to maximize the social
welfare subject to its budget constraint. A naive approach in which
the government chooses the pair that maximizes the overall grade is
clearly manipulable (there may be settings where an agent will
falsely down-grade his 2nd choice to the third place in order to
ensure his first choice makes it). An alternative methodology is for
the government to choose a pair randomly, where the probability
assigned to each pair is an increasing function of its induced
welfare (the specific nature of the function will be made clear in
the sequel). In addition, with a vanishing probability, a random pair will be chosen and in that case each agent will assigned her preferred program according to her announcement.

It turns out that this scheme is not manipulable and agents' optimal
strategy is to report truthfully. If the population is large enough
then the probability of choosing the truly optimal pair can be made
arbitrarily close to one. \eexa

\subsection{Related Work}

\paragraph{Virtual implementation.} The most closely related body of work is the literature on `virtual implementation' with incomplete information, derived from earlier work on virtual implementation with complete information which was
initiated by Matsushima \cite{Matsushima88} and Abreu and Sen
\cite{Abreu and Sen 90}. A social choice function is {\em virtually
implementable} if for any $\epsilon >0$ there exists a mechanism
which equilibria result in outcomes that $\epsilon$-approximate the
function. Results due to Abreu and
Matsushima \cite{AM92}, Duggan \cite{Duggan97} and Serrano and Vohra \cite{SerranoVohra01,
SerranoVohra05} provide necessary and sufficient conditions for
functions to be virtually implementable in various environments with
private information. A common thread throughout the results on virtual implementation under incomplete information is the
incentive compatibility requirement over the social choice function,
in addition to some form of type diversity. Compared with our
contribution the above mentioned work provides positive results in
environments with small populations, whereas we require large
populations in order to have a meaningful approximation. On the
other hand, the solution concepts we focus on are ex-post Nash
equilibrium, undominated strategies, and strict dominance (for the private values setting),
compared with iterated deletion of
dominated strategies or Bayes-Nash equilibria, provided in the above
mentioned papers. In addition, the virtual implementation results apply to functions that are incentive compatible from the outset, whereas our technique applies to arbitrary objective
functions. In both cases the mechanisms
proposed do not require transfers but do require some kind of
type diversity.

\paragraph{Influence and Approximate Efficiency.} The basic driving force underlying our construction
is ensuring that each agent has a vanishing influence on the outcome
of the mechanism as the population grows. In the limit, if players
are non-influential, then they might as well be truthful. This idea
is not new and has been used by various authors to provide
mechanisms that approximate efficient outcomes when the population
of players is large. Some examples of work that hinge on a similar
principle for large, yet finite populations, are Swinkels
\cite{Swinkels} who studies auctions, Satterthwaite and Williams
\cite{SW} and Rustichini, Satterthwaite and Williams \cite{RSW} who
study double auctions,  and  Al-Najjar and Smorodinsky \cite{AS2007}
who study an exchange market. The same principle is even more
enhanced in models with a continuum of players, where each agent has
no influence on the joint outcome (e.g., Roberts and Postlewaite
\cite{RP76} who study an exchange economy). The mechanisms provided
in these papers are designed for maximizing the sum of agents'
valuations, and provide no value for alternative objective
functions. In contrast, our results hold a for a wide range of
objective functions and are generic in nature. Interestingly, a
similar argument, hinging on players' lack of influence, is
instrumental to show inefficiency in large population models (for
example,  Mailath and Postlewaite \cite{MP90} demonstrate
`free-riding' in the context of public goods, which eventually leads
to inefficiency).

A formal statement of `influence' in an abstract setting appears in Levine and Pesendorfer \cite{LP95} and Al-Najjar and
Smorodinsky \cite{AS2000}. Beyond the formalization of influence these works provide bounds on aggregate measures of influence such as the average influence or on the number of influential agents. McLean and Postlewaite
\cite{MP2002} introduce the notion of informational smallness,
formalizing settings where one player's information is insignificant
with respect to the aggregated information.

\paragraph{Differential Privacy.}
The notion of {\em differential privacy}, recently introduced by Dwork, McSherry, Nissim and Smith~\cite{DMNS06}, captures a measure of (lack of) privacy by the impact of a single agent's input on the outcome of a joint computation. A small impact suggests that the agent's privacy cannot be significantly jeopardized. In the limit, if an agent has no impact then nothing can be learned about the agent from the outcome of the computation. More accurately, differential privacy
stipulates that the influence of any contributor to the computation
is bounded in a very strict sense: any change in the input
contributed by an individual translates to at most a near-one
multiplicative factor in the probability distribution over the set
of outcomes.%
\footnote{The measure of `impact' underlying differential privacy is the analog of `influence' a-la Levine and Pesendorfer~\cite{LP95} and Al-Najjar and
Smorodinsky~\cite{AS2000} in a non-Bayesian framework, with worst-case considerations.}

The scope of computations that were shown to be
computed in a differentially private manner has grown significantly
since the introduction of the concept and the reader is referred to
Dwork \cite{Dwork09} for a recent survey.

McSherry and Talwar~\cite{MT07} establish an inspiring connection
between differential privacy and mechanism design, where differential privacy is used as a tool for constructing efficient mechanisms. They observe that
participants (players) that contribute private information to
$\epsilon$-differentially private computations have limited
influence on the outcome of the computation, and hence have a
limited incentive to lie, even if their utility is derived from the
joint outcome.
Consequently, truth-telling is approximately dominant in mechanisms that are
$\epsilon$-differentially private, regardless of the agent utility functions.%
\footnote{Schummer \cite{Schummer2004} also studies approximately
dominant strategies, in the context of exchange economies.}
McSherry and Talwar introduce the exponential mechanism as a generic
$\epsilon$-differentially private mechanism. In addition, they show
that whenever agents are truthful the exponential mechanism chooses
a social alternative which almost optimizes the objective function.
They go on and demonstrate the power of this mechanism in the
context of Unlimited Supply Auctions, Attribute Auctions, and
Constrained pricing.

The contribution of McSherry and Talwar leaves
much to be desired in terms of mechanism design:
(1) It is not clear how to set the value of $\epsilon$. Lower values of $\epsilon$ imply higher compatibility with incentives, on the one hand, but deteriorate the approximation results on the other hand. The model and results of McSherry and Talwar do not provide a framework for analyzing these countervailing forces.
(2) Truth telling is {\em approximately} dominant, but, in fact, in the mechanisms they design {\bf all} strategies are approximately dominant,
which suggests that truth telling may have no intrinsic advantage over
any other strategy in their mechanism. (3) Furthermore, one can
demonstrate that misreporting one's private information can actually
dominate other strategies, truth-telling included. To make things
worse, such dominant strategies may lead to inferior results for the
social planner. This is demonstrated in Example~\ref{example1}, in
the context of monopoly pricing.

\paragraph{Facility Location.}\label{sec5}

One of the concrete examples we investigate is the optimal location of facilities.
The facility location problem has already been tackled in the
context of approximate mechanism design without money,
and turned out to lead to interesting challenges. While the single facility location problem
exhibits preferences that are single-peaked and can be solved optimally by selecting the
median declaration, the 2-facility problem turns out to be non-trivial. Most recently
Wang et al \cite{facility10} introduce a randomized $4$-(multiplicative)
approximation  truthful mechanism for the $2$ facility location problem.
The techniques introduced here provide much better approximations - in particular we provide an
additive $\tilde O(n^{-1/3})$ approximation to the average optimal distance between the agents and the facilities.%
\footnote{The notation $\tilde O({n^{-1}})$ is used to denote convergence to zero at a
rate $\frac{\ln(n)}{n}$. Compared with $O({n^{-1}})$ which denotes convergence to zero at a rate $\frac{1}{n}$ }

Following our formalization of {\em reactions} and of {\em imposition} and its applicability to facility location, Fotakis and Tzamos \cite{Fotakis_Tzamos_10} provide `imposing' versions of previously known mechanisms to improve implementation accuracy. They provide constant multiplicative approximation or logarithmic multiplicative approximation, albeit with fully imposing mechanisms.

\paragraph{Non discriminatory Pricing of Digital Goods.} Another concrete setting where we demonstrate
our generic results is a pricing application, where a monopolist
sets a single price for goods with zero marginal costs (``digital
goods") in order to maximize revenues. We consider environments
where the potential buyers have {\em interdependent valuations} for the
good. Pricing mechanisms for the {\em private values} case have been
studied by Goldberg et al \cite{Goldberg_etal} and Balcan et al
\cite{Balcan05}. They consider settings where agents' valuation are not necessarily restricted to a finite set and achieve $O(\frac{1}{\sqrt{n}})$-implementation (where $n$ is the
population size). Whereas our mechanism provides a similar bound it is limited to settings with finitely many possible prices. However, it is derived from
general principles and therefore more robust. In addition, our
mechanism is applicable beyond the private values' setting.


\section{Model}

\subsection{The Environment}

Let $N$ denote a set of $n$ agents, $S$ denotes a {\em finite} set
of social alternatives and
$T_i$, $i=1,\dots,n$, is a finite type
space for agent $i$. We denote by $T =  \times_{i=1}^n T_i$ the set
of type tuples and write $T_{-i} = \times_{j \not = i}T_j$ with
generic element $t_{-i}$. Agent $i$'s type, $t_i \in T_i$, is her
private information.
Let $R_i$ be the set of reactions available to $i$. Typically, once a social alternative,
$s \in S$, is determined agents choose a reaction $r_i \in R_i$. The
utility of an agent $i$ is therefore a function of the vector of
types, the chosen social alternative and the chosen reaction.
Formally, $u_i:T \times S \times R_i  \to [0,1]$.%
\footnote{Utilities are assumed to be bounded in the unit interval. This is without loss of generality, as long as there is some uniform bound on the utility.}
A tuple $(T,S,R,u)$, where $R=\times_{i=1}^n R_i$ and $u=(u_1,\ldots,u_n)$, is called an {\em environment}.
We will use $r_i(t,s)$ to denote an arbitrary optimal reaction for agent $i$ (i.e., $r_i(t,s)$ is an arbitrary function which image is in the set $\argmax_{r_i \in R_i} u_i(t,s,r_i)$).

We say that an agent has {\em private reactions} if her optimal reaction
of $i$ depends only only on her type and the social alternative.
Formally, agent $i$ has private reactions if $\argmax_{r_i \in
R_i} u_i((t_i,t_{-i}),s,r_i) = \argmax_{r_i \in R_i}
u_i((t_i,t'_{-i}),s,r_i)$, for all $s, i, t_i,t_{-i}$ and $t'_{-i}$. To emphasize that $r_i(t,s)$ does not depend on $t_{-i}$ we will use in this case the notation $r_i(t_i,s)$ to denote an arbitrary optimal reaction for agent $i$.
We say that an agent has {\em private values} if she has private reactions and furthermore her utility depends only on her type, social alternative and reaction, i.e., $u_i((t_i,t_{-i}),s,r_i) = u_i((t_i,t'_{-i}),s,r_i)$ for all $s, i, t_i,t_{-i}$ and $t'_{-i}$.
In this case we will use the notation $u_i(t_i,s,r_i)$ to denote the agent's utility, to emphasize that it does not depend on $t_{-i}$. In the  more general setting, where the utility $u_i$ and the optimal reaction $r_i$ may depend on $t_{-i}$, we say that agents have {\em interdependent values}.

An environment is {\em non-trivial} if for any pair of
types there exists a social alternative for which the optimal
reactions are distinct.  Formally, $\forall i$, $t_i \not = \hat t_i
\in T_i$ and  $t_{-i}$ there exists $s \in S$, denoted  $s(t_i,\hat
t_i,t_{-i})$, such that $\argmax_{r_i \in R_i}
u_i((t_i,t_{-i}),s,r_i) \ \cap \ \argmax_{r_i \in R_i} u_i((\hat
t_i,t_{-i}), s,r_i) = \emptyset $.
We say that  $s(t_i,\hat t_i,t_{-i})$ {\em separates} between  $t_i$ and $\hat t_i$ at $t_{-i}$. A set of social alternatives, $\tilde S \subset S$ is called {\em separating} if for any $i$ and $t_i \not = \hat t_i$ and $t_{-i}$, there exists some $s(t_i,\hat t_i,t_{-i}) \in \tilde S$ that separates between  $t_i$ and $\hat t_i$ at $t_{-i}$.

\subsection{The Objective Function}

A social planner, not knowing the vector of types, wants to maximize an arbitrary {\em objective function}
(sometimes termed {\em social welfare function}), $F:T \times S \to [0,1]$.%
\footnote{In fact, one can consider objective functions of the form
$F:T \times S \times R \to [0,1]$. Our results go through if for any
$t$ and $s$ and any $i$ and $r_{-i}$ the functions
$F(t,s,(r_{-i},\cdot)):R_i \to [0,1]$ and $u_i(t,s,\cdot):R_i
\to[0,1]$ are co-monotonic. In words, as long as the objective
function's outcome (weakly) increases whenever a change in reaction increases an agent's utility.} We focus our
attention on a class of functions for which individual agents have a
diminishing impact, as the population size grows:

\bdefi[Sensitivity]
The objective function $F:T \times S \to [0,1]$  is {\em $d$-sensitive}
if $\forall i, t_i \not = \hat t_i, t_{-i}$ and $s \in S$, $\ |F((t_i,t_{-i}),s) - F((\hat t_i, t_{-i}),s) | \le {d\over n}$,
where $n$ is the population size.%
\footnote{
In the definition of sensitivity one can replace the constant $d$ with a function $d=d(n)$ that depends on the population size. Our go through for the more general case as long as $\lim_{n\to\infty} \frac{d(n)}{n}=0$.}
\edefi

Note that this definition refers to unilateral changes in
announcements, while keeping the social alternative fixed. In
particular  $d$-sensitivity does not exclude the possibility of a
radical change in the optimal social alternative as a result of
unilateral deviations, which, in turn, can radically change the
utility of the player. Thus, this definition is mute in the context
of the influence of an agent on her own utility.

One commonly used objective function which is $1$-sensitive is the average utility,
$$F(t,s)={\sum_i u_i(t,s,r_i(t,s)) \over n}.$$
Note that a $d$-sensitive function eliminates
situations where any single agent has an overwhelming impact on the value of the objective function,
for a fixed social alternative $s$.  In fact, if an objective function is not $d$-sensitive, for any $d$,
then in a large population this function could be susceptible to minor faults in the system
(e.g., noisy communication channels).%
 \footnote{An example of a function that is not $d$-sensitive, for any $d$, is the following:
 set $F=1$ ($F=0$) if there is an even number of agents which utility exceeds some threshold and the social alternative is $A$ ($B$),
 and $F=0$ ($F=1$) otherwise.}

\subsection{Mechanisms}

Denote by ${\cal R}_i = 2^{R_i}\setminus \{\emptyset\}$ the set of all subsets of $R_i$, except for the empty set, and let ${\cal R} = \times_i {\cal R}_i$.

A (direct) mechanism randomly chooses, for any vector of inputs $t$ a social alternative, and for each agent $i$  a subset of available reactions. Formally:
\bdefi[Mechanism]
A (direct) {\em mechanism} is a function $M:T \to \Delta(S \times {\cal R})$.
\edefi

In addition, the mechanism discloses the vector of agents' announcements, and agents can use this information to choose a reaction.\footnote{If, however, all agents have private reactions then this information is useless to the agents and we do not require such a public disclosure of the agents' announcements.}

We denote by $M_S(t)$ the marginal distribution of $M(t)$ on $S$ and by $M_i(t)$ the marginal distribution on ${\cal R}_i$. We say that the mechanism $M$ is {\em non-imposing} if $M_i(t)(R_i)=1$. That is, the probability assigned to the grand set of reactions is one, for all $i$ and $t \in T$. Put differently, the mechanism never restricts the set of available reactions.  $M$ is {\em $\epsilon$-imposing} if $M_i(t)(R_i) \ge 1-\epsilon$ for all $i$ and $t \in T$. In words, with probability exceeding $1-\epsilon$ the mechanism imposes no restrictions.

\subsection{Strategies and Solution Concepts}

A mechanism induces the following game with incomplete information. In the first phase agents announce their types simultaneously to the mechanism. Then the mechanism chooses a social alternative and a subset of reactions for each agent.
In the second stage of the game each agent, knowing the strategy tuple of all agents, the vector of announced types, the social alternative and her available set of reactions, must choose one such reaction. Let $W_i:T_i \to T_i$ denote the announcement of agent $i$, given his type and let $W=(W_i)_{i=1}^n$. Upon the announcement of the social alternative $s$, the vector of opponents' announcements, $t_{-i}$ and a subset of reactions, $\hat R_i \subset R_i$, the rational agent will choose an arbitrary optimal reaction, $r_i((t_i,W^{-1}_{-i}(t_{-i})),s, \hat R_i)$, where $W^{-1}_{-i}(t_{-i})$ denotes the pre-image of $W_{-i}$ at the vector of announcements $t_{-i}$.%
\footnote{We slightly abuse notation as $W^{-1}_{-i}(t_{-i})$ may not be a singleton but a subset of type vectors, in which case the optimal reaction is not well defined. More accurate notation must involve considering another primitive to the model - the prior belief of $i$ over $T_{-i}$. With such a prior $r_i((t_i,W^{-1}_{-i}(t_{-i})),s, R_i)$ denotes the reaction in $\hat R_i$ that maximizes the expected utility with respect to the prior belief, conditional on the subset  $W^{-1}_{-i}(t_{-i})$.}

Thus, given a mechanism and a vector of announcement functions, $(W_i)_{i=1}^n$,  the agents' reaction are uniquely defined. Therefore, we can view $(W_i)_{i=1}^n$ as the agents' strategies, without an explicit reference to the choice of reactions. Given a vector of types, $t$, and a strategy tuple $W$, the mechanism $M$ induces a probability distribution, $M(W(t))$ over the set of social alternatives and reaction tuples. The expected utility of $i$, at a vector of types $t$, is $E_{M(W(t))}u_i(t,s,r_i)$, where $r_i$ is short-writing for the optimal reaction, which itself is determined by $M$ and $W$. In fact, hereinafter we suppress the reference to the reactions in our notations and write
$E_{M(W(t))}u_i(t,s)$ instead of $E_{M(W(t))}u_i(t,s,r_i)$.

A strategy $W_i$ is {\em dominant} for the mechanism $M$ if for any
vector of types $t\in T$, any alternative strategy $\hat W_i$ of $i$
and any strategy profile $\bar W_{-i}$ of $i$'s opponents
\begin{equation}\label{eq:dominance}
E_{M((W_i(t_i),\bar W_{-i}(t_{-i})))} u_i(t,s) \ge E_{M((\hat
W_i(t_i),\bar W_{-i}(t_{-i})))} u_i(t,s).
\end{equation}

In words, $W_i$ is a
strategy that maximizes the expected payoff of $i$ for any vector of
types and any strategy used by her opponents. If for all $i$ the
strategy $W_i(t_i)=t_i$ is dominant
then $M$ is called {\it truthful} (or {\it strategyproof}).%
\footnote{Note we do not require a strong inequality to hold on any instance.}

A strategy $W_i$ is {\em strictly dominant} if it is dominant and furthermore whenever $W(t_i) \not = \hat W(t_i)$ then a strong inequality holds in Equation~(\ref{eq:dominance}). If $W_i(t_i)=t_i$ is strictly dominant for all $i$ then $M$ is {\it strictly truthful}.

A strategy $W_i$ is {\em dominated} for the mechanism $M$ if there exists an alternative strategy $\hat W_i$, such that for any vector of types $t\in T$, and any strategy profile $\bar W_{-i}$ of $i$'s opponents, the following holds:
$E_{M((W_i(t_i),\bar W_{-i}(t_{-i})))} u_i(t,s) \le E_{M((\hat W_i(t_i),\bar W_{-i}(t_{-i})))} u_i(t,s),$ with a strong inequality holding for at least one type vector $t$.

Finally, a strategy tuple $W$ is an {\em ex-post Nash Equilibrium} if for all $i$ and $t \in T$ and for any strategy $\hat W_i$ of player $i$,
$E_{M(W(t))} u_i(t,s) \ge E_{M((\hat W_i(t_i),W_{-i}(t_{-i})))} u_i(t,s)$.
If $\{W_i(t_i)=t_i\}_{i=1}^n$ is an ex-post Nash equilibrium then $M$ is {\it ex-post Nash truthful}.%

\subsection{Implementation}

Given a vector of types, $t$, the expected value of the objective function, $F$, at the strategy tuple $W$ is $E_{M(W(t))}[F(t,s)]$.

\bdefi[$\beta$-implementation]
We say that the mechanism $M$ {\em $\beta$-implements $F$ in (strictly) dominant strategies}, for $\beta > 0$,
if for any (strictly) dominant strategy tuple, $W$, for any $t \in T$, $\ E_{M(W(t))}[F(t,s)] \ge max_{s \in S} F(t,s)-\beta$.

A mechanism $M$ {\em $\beta$-implements $F$ in an ex-post Nash equilibrium} if for some ex-post Nash equilibrium
strategy tuple, $W$, for any $t \in T$, $\ E_{M(W(t))}[F(t,s)] \ge max_{s \in S} F(t,s)-\beta$.

A mechanism $M$ {\em $\beta$-implements $F$ in undominated strategies} if for any tuple of strategies, $W$,
that are not dominated and for any $t \in T$, $\ E_{M(W(t))}[F(t,s)] \ge max_{s \in S} F(t,s)-\beta$.
\edefi

\paragraph{Main Theorem  (informal statement):} For any $d$-sensitive function $F$ and $1 > \beta > 0$
there exists a number $n_0$ and a mechanism $M$ which $\beta$-implements $F$ in an ex-post Nash equilibrium,
whenever the population has more than $n_0$ agents.
If, in addition, reactions are private then $M$ $\beta$-implements $F$ in strictly dominant strategies.

\section{A Framework of Approximate Implementation}

In this section we present a general scheme for implementing
arbitrary objective functions in large societies. The convergence
rate we demonstrate is of an order of magnitude of $\sqrt{\frac{\ln
(n)}{n}}$. Our scheme involves a lottery between two mechanisms: (1)
The {\em Exponential Mechanism}, a non-imposing differentially-private mechanism that randomly
selects a social alternative, $s$. The probability of choosing $s$
is proportional to (a exponent of) the value it induces on $F$; and
(2) The {\em Commitment Mechanism}, where imposition is used to commit agents to take a reaction that complies with their announced type.

\subsection{The Exponential Mechanism and Differential Privacy}

Consider the following non-imposing mechanism, which we refer to as
the {\em Exponential Mechanism}, originally introduced by McSherry and
Talwar \cite{MT07}:
$$ M^{\epsilon}(t)(s) =\frac{e^{n\epsilon F(t,s)}}{\sum_{\bar s \in S} e^{n\epsilon F(t,\bar s)}}.$$

The Exponential mechanism has two notable properties, as we show
below: It provides $\epsilon$-differential privacy, i.e., for all $i$ it is insensitive to a change in $t_i$. And, it chooses $s$ that almost maximizes $F(t,s)$.

We follow Dwork et al~\cite{DMNS06} and define:

\bdefi\label{privacy}[$\epsilon$-differential privacy] A mechanism, $M$, provides
{\em $\epsilon$-differential privacy} if it is non-imposing and for
any $s \in S$, any pair of type vectors $t, \hat t \in T$,
which differ only on a single coordinate, $M(t)(s) \leq e^\epsilon \cdot M(\hat t)(s)$.%
\footnote{For non discrete sets of alternatives the definition requires that
$\frac{M(t)(\hat S)}{M(\hat t)(\hat S)} \leq e^\epsilon \ \  \forall \hat S \subset S $.}%
\edefi
In words, a mechanism preserves  $\epsilon$-differential privacy if, for any vector of
announcements, a unilateral deviation changes the probabilities assigned to any social
choice $s \in S$ by a (multiplicative) factor of $e^\epsilon$, which approaches $1$ as $\epsilon$ approaches zero.%
\footnote{The motivation underlying this definition of $\epsilon$-differential
privacy is that if a single agent's input to a database changes then a query on that database would result in
(distributionally) similar results.
This, in return, suggests that it is difficult to learn new information about the agent from the query, thus preserving her privacy.}

\blem[McSherry and Talwar \cite{MT07}]\label{lemma1}
If $F$ is $d$-sensitive then $M^{\frac{\epsilon}{2d}}(t)$ preserves $\epsilon$-differential privacy.
\elem

The proof is simple, and is provided for completeness:

{\bf Proof}: Let $t$ and $\hat t$ be or two type vectors that differ on a single coordinate. Then for any $s \in S$, $\ F(t,s)-\frac{d}{n} \leq F(\hat t,s) \leq F(t,s)+\frac{d}{n}$, hence,
$$
\frac{M^{\frac{\epsilon}{2d}}(t)(s)}{M^{\frac{\epsilon}{2d}}(\hat t)(s)} = \frac{ \frac{e^{\frac{n\epsilon F(t,s)}{2d}}}{\sum_{\bar s \in S} e^{\frac{n\epsilon F(t,\bar s)}{2d}} } }
{
\frac{e^{\frac{n\epsilon F(\hat t,s)}{2d}}}{\sum_{\bar s \in S} e^{\frac{n\epsilon F(\hat t,\bar s)}{2d}} }} \le
\frac {\frac{e^{\frac{n\epsilon F(t,s)}{2d}}}{\sum_{\bar s \in S} e^{\frac{n\epsilon F(t,\bar s)}{2d}} } }
{\frac{e^{\frac{n\epsilon (F(t,s)-\frac{d}{n})}{2d}}}{\sum_{\bar s \in S} e^{\frac{n\epsilon (F(t,\bar s)+\frac{d}{n})}{2d}} } } = e^{\epsilon}.
$$

{\bf QED}

The appeal of mechanisms that provide $\epsilon$-differential privacy is that they induce near indifference among all strategies, in the following sense:

\blem\label{lemma2}
If $M$ is non-imposing and provides $\epsilon$-differential privacy, for some $\epsilon <1$, then for any agent $i$, any type tuple $t$, any strategy tuple $W$, and any alternative strategy for $i$, $\hat W_i$ the following holds:
$$|E_{M(W(t))} [u_i(t,s)] - E_{M(\hat W_i(t_i),W_{-i}(t_{-i}))} [u_i(t,s)] | <  2\epsilon.$$
\elem

The proof is simple, and is provided for completeness:

{\bf Proof}:
Let $W$ and $\hat W$ be two strategy vectors that differ the $i$'th coordinate. Then for every $t\in T$, $s\in S$, $r_i \in R_i$ and $u_i: T \times S \times R_i \rightarrow [0,1]$ we have
\begin{eqnarray*}
 E_{M(W(t))} [u_i(t,s)] & = & \sum_{s\in S} M(W(t))(s) \cdot u_i(t,s) \\
& \leq & \sum_{s\in S} e^{\epsilon} \cdot M(\hat W_i(t_i),W_{-i}(t_{-i}))(s) \cdot u_i(t,s) \\
& = & e^{\epsilon} \cdot E_{\hat W_i(t_i),W_{-i}(t_{-i}))} [u_i(t,s)],
\end{eqnarray*}
where the inequality follows since $M$ provides $\epsilon$-differential privacy, and $u_i$ is non-negative. A similar analysis gives
$$ E_{\hat W_i(t_i),W_{-i}(t_{-i}))} [u_i(t,s)] \leq e^{\epsilon}\cdot E_{M(W(t))} [u_i(t,s)].$$
Hence we get:
\begin{eqnarray*}
E_{M(W(t))} [u_i(t,s)] - E_{M(\hat W_i(t_i),W_{-i}(t_{-i}))} [u_i(t,s)] &\leq &(e^{\epsilon}-1) \cdot E_{M(\hat W_i(t_i),W_{-i}(t_{-i}))} [u_i(t,s)] \\
& \leq & e^{\epsilon}-1,
\end{eqnarray*}
where the last inequality holds because $u_i$ returns a values in
$[0,1]$. Similarly,
$$E_{M(\hat W_i(t_i),W_{-i}(t_{-i}))} [u_i(t,s)] - E_{M(W(t))} [u_i(t,s)] \leq e^{\epsilon}-1.$$

To conclude the lemma, note that $(e^{\epsilon}-1) \le 2\epsilon$
for $\ 0 \leq \epsilon \leq 1$.

{\bf QED}

McSherry and Talwar \cite{MT07} note in particular that in the case
of private values truthfulness is $2\epsilon$-dominant, which is an
immediate corollary of Lemma \ref{lemma2}. They combine this with
the following observation to conclude that exponential mechanisms
approximately implement $F$ in $\epsilon$- dominant strategies:

\blem[McSherry and Talwar \cite{MT07}]\label{lemma3} Let $F:T^n
\times S \to [0,1]$ be an arbitrary $d$-sensitive objective function
and $n > \frac{e2d}{\epsilon |S|}$. Then for any $\ t$,
$E_{M^{\frac{\epsilon}{2d}}(t)}[F(t,s)] \ge \max_s F(t,s) -
\frac{4d}{n\epsilon}\ln\left(\frac{n\epsilon|S|}{2d}\right)$. \elem

{\bf Proof}: Let $\delta = \frac{2d}{n\epsilon}\ln
\left(\frac{n\epsilon|S|}{2d}\right)$.
As $n > \frac{e2d}{\epsilon
|S|}$ we conclude that $\ln\left(\frac{n\epsilon|S|}{2d}\right)>\ln e>0$
and, in particular, $\delta > 0$.

Fix a vector of types, $t$ and denote by
$\hat S = \{\hat s \in S :
F(t, \hat s) < \max_s F(t,s) - \delta \}$. For any $\hat s \in \hat
S$ the following holds:
$$M^{\frac{\epsilon}{2d}}(t)(\hat s) = \frac{e^{\frac{n\epsilon F(t, \hat s)}{2d}}}{\sum_{s'\in S} e^{\frac{n\epsilon F(t, s')}{2d}}} \leq
\frac{e^{\frac{n\epsilon (\max_s F(t,s)-\delta)}{2d}}}{e^{\frac{n\epsilon \max_s F(t,s)}{2d}}} = e^{-\frac{n\epsilon}{2d}\delta}.$$

Therefore,
$M^{\frac{\epsilon}{2d}}(t)(\hat S) = \sum_{\hat s \in \hat S } M^{\frac{\epsilon}{2d}}(t)(\hat s) \le |\hat S| e^{-\frac{n\epsilon}{2d}\delta} \le  |S| e^{-\frac{n\epsilon}{2d}\delta}$.
Which, in turn, implies:
$$E_{M^{\frac{\epsilon}{2d}}(t)}[F(t,s)] \geq (\max_s F(t,s)-\delta) (1-|S| e^{-\frac{n\epsilon}{2d}\delta})  \geq
\max_s F(t,s) - \delta - |S| e^{-\frac{n\epsilon}{2d}\delta}.$$

Substituting for $\delta$ we get that
$$E_{M^{\frac{\epsilon}{2d}}(t)}[F(t,s)]  \geq \max_s F(t,s) -
 \frac{2d}{n\epsilon}\ln \left(\frac{n\epsilon|S|}{2d}\right) -
 \frac{2d}{n\epsilon}.$$

In addition, $n > \frac{e2d}{\epsilon |S|}$ which implies  $\ln \left(\frac{n\epsilon|S|}{2d}\right) > \ln (e)=1$,
and hence $\frac{2d}{n\epsilon} \leq \frac{2d}{n\epsilon}\ln \left(\frac{n\epsilon|S|}{2d}\right)$.
Plugging this into the previous inequality yields
$E_{M^{\frac{\epsilon}{2d}}(t)}[F(t,s)]  \geq \max_s F(t,s) - \frac{4d}{n\epsilon}\ln \left(\frac{n\epsilon|S|}{2d}\right)$ as desired.

{\bf QED}

Note that $\lim_{n \to \infty}\frac{4d}{n\epsilon}\ln\left(\frac{n\epsilon|S|}{2d}\right) = 0$
whenever the parameters $d,\epsilon$ and $|S|$ are held fixed.%
\footnote{This limit also approaches zero if $d,\epsilon,|S|$ depend
on $n$, as long as $d/\epsilon$ is sublinear in $n$ and $|S|$ is
subexponential in $n$.} Therefore, the exponential mechanism is
almost optimal for a large and truthful population.

\paragraph{Remark:} There are other mechanisms which exhibit similar properties to
those of the Exponential Mechanism, namely `almost indifference' and
`approximate optimality'. The literature on differential privacy is
rich in techniques for establishing mechanisms with such properties.
Some techniques for converting computations into
$\epsilon$-differentially private computations without jeopardizing
the accuracy too much are the addition of noise calibrated to global
sensitivity by Dwork et al.~\cite{DMNS06}, the addition of noise
calibrated to smooth sensitivity and the sample and aggregate
framework by Nissim et al.~\cite{NRS07}. The reader is further
referred to the recent survey of Dwork~\cite{Dwork09}.
Any of these mechanisms can replace the exponential mechanism in the following analysis.

\subsection{The Commitment Mechanism}

We now consider an imposing mechanism that chooses $s\in S$ randomly, while ignoring agents' announcements. Once $s$ is chosen the mechanism restricts the allowable reactions for $i$ to those that are optimal assuming all agents are truthful. Formally, if $s$ is chosen according to the probability distribution $P$, let $M^P$ denote the following mechanism:  $$M_S^P(t)(s) = P(s)\quad\mbox{\rm and}\quad M_i^P(t)(r_i(t,s)) | s) = 1.$$ Players do not influence the choice of $s$ in $M^P$
and so they are (weakly) better off being truthful.

We define the {\em gap} of the environment, $\gamma = g(T,S,A,u)$, as:
$$\gamma = g(T,S,A,u) = \min_{i,t_i \not = b_i,t_{-i}} \max_{s\in S} \left( u_i(t,s,r_i(t,s)) - u_i(t,s,r_i((b_i,t_{-i}),s)) \right).$$

In words, $\gamma$ is a lower bound for the loss incurred by misreporting in case of an adversarial choice of $s \in S$. In non-trivial environments $\gamma >0$. We say the a distribution $P$ is {\em separating} if there exists a separating set $\tilde S \subset S$ such that $P(\tilde s)>0$ for all $\tilde s \in \tilde S$.
In this case we also say that $M^P$ is a separating mechanism. In particular let $\tilde p = \min_{s \in \tilde S}P(s)$. Clearly one can choose $P$ such that $\tilde p \ge \frac{1}{|S|} $. The following is straightforward:

\blem\label{lemma4}
If the environment $(T,S,A,u)$ is non-trivial and $P$ is a separating distribution over $S$ then $\forall b_i\not=t_i, t_{-i}$,
$$E_{M^P(t_i,t_{-i})}[u_i(t,s,r_i(t,s) )] \ge
E_{M^P(b_i,t_{-i})}[u_i(t,s, r_i((b_i,t_{-i}),s))] + \tilde p \gamma \ .$$

If, in addition, reactions are private, then for any $i$, $b_i\not=t_i, \ t_{-i}$ and $b_{-i}$:
$$E_{M^P(t_i,b_{-i})}[u_i(t,s,r_i(t_i,s) )] \ge
E_{M^P(b_i,b_{-i})}[u_i(t,s, r_i(b_i,s))] + \tilde p \gamma. $$
\elem

{\bf Proof}: For any pair $b_i \not = t_i$ and for any $s \in S$ $
u_i(t,s,r_i(t_i,s)) \ge u_i(t,s,r_i(b_i,s))$. In addition, there
exists some $\hat s = s(t_i,b_i)$, satisfying $P(\hat s) \ge \tilde
p$, for which $ u_i(t,\hat s,r_i(t_i,\hat s)) \ge u_i(t,\hat
s,r_i(b_i,\hat s)) + \gamma$. Therefore, for any $\ i$, $\ b_i \not
= t_i \in T_i$ and for any $t_{-i}$,
$E_{M^P(t_i,t_{-i})}[u_i(t,s,r_i(t,s) )] \ge
E_{M^P(b_i,t_{-i})}[u_i(t,s, r_i((b_i,t_{-i}),s))] + \tilde p
\gamma \ $, as claimed.

Recall that if reactions are private then $r_i(t,s) = r_i(t_i,s)$,
namely the optimal reaction of an agent, given some social
alternative $s$, depends only on the agent's type. Therefore we
derive the result for private reactions by replacing
$r_i((t_i,t_{-i}),s)$ with $r_i(t_i,s)$ on the left hand
side of the last inequality and $r_i((b_i,t_{-i}),s)$ with
$r_i(b_i,s)$ on the right hand side.

QED

The following is an immediate corollary:

\bcor\label{cor1}
If the environment $(T,S,A,u)$ is non-trivial and $P$ is a separating distribution over $S$ then
\begin{enumerate}
\item
Truthfulness is an ex-post Nash equilibrium of $M^P$.
\item
If agent $i$ has private reactions then truthfulness is a strictly dominant strategy for $i$ in $M^P$.
\end{enumerate}
\ecor

An alternative natural imposing mechanism is that of a random
dictator, where a random agent is chosen to dictate the social
outcome. Similarly, agents will be truthful in such a mechanism.
However, the loss from misreporting can only be bounded below by
$\frac{\gamma}{n}$, whereas the commitment mechanism gives a lower bound of $\gamma\tilde p \ge
\frac{\gamma}{|S|}$, which is independent of the population size.

\subsection{A Generic and Nearly Optimal Mechanism}

Fix a non-trivial environment $(T,S,A,u)$ with a gap $\gamma$, separating set $\tilde S$, a $d$-sensitive objective function $F$ and a separating commitment mechanism, $M^P$, with $\tilde p = \min_{s \in \tilde S}P(s)$.

Set $\bar M_q^\epsilon(t) = (1-q)M^{\frac{\epsilon}{2d}}(t) + q M^P(t)$.

\bthm\label{theorem1}
If $q \tilde p \gamma \ge 2\epsilon$ then the mechanism $\bar M_q^\epsilon$ is ex-post Nash truthful. Furthermore, if agents have private reactions then $\bar M_q^\epsilon$ is strictly truthful.
\ethm

{\bf Proof}: Follows immediately from Lemmas ~\ref{lemma2} (set $W(t_i)=t_i \ $) and ~\ref{lemma4}.

{\bf QED}

Set the parameters of the mechanism $\bar M_q^\epsilon(t)$ as follows:
\begin{itemize}
\item
$\epsilon = \sqrt{\frac{\tilde p \gamma d }{n}}\sqrt{\ln \left(\frac{n\tilde p\gamma|S|}{2d}\right)}$.
\item
$q = \frac{2\epsilon}{\tilde p \gamma}$.
\end{itemize}
and consider populations of size $n>n_0$, where $n_0$ is the minimal
integer satisfying $n_0 \ge \max\{ \frac{8 d}{\tilde p \gamma} \ln
\left(\frac{\tilde p\gamma|S|}{2d}\right), \frac{4e^2d}{\tilde p
\gamma |S|} \}$ and $\frac{n_0}{\ln(n_0)} > \frac{8 d}{\tilde p
\gamma}$.

\blem\label{lemma5}
If $n>n_0$ then
\begin{enumerate}
\item
$q=\frac{2\epsilon}{\tilde p \gamma} <1$.
\item
$ \epsilon < \tilde p \gamma$.
\item
$n > \frac{2ed}{\epsilon|S|}$.
\end{enumerate}
\elem

{\bf Proof}:
Part (1):
$ \frac{n}{\ln(n)} >\frac{n_0}{\ln(n_0)} \ge \frac{8 d}{\tilde p \gamma}$
which implies
$ n > \frac{8 d}{\tilde p \gamma}\ln(n)$. In addition,
$ n >n_0 >  \frac{8 d}{\tilde p \gamma} \ln \left(\frac{\tilde p \gamma|S|}{2d}\right)$. Therefore
$ n >\frac{4 d}{\tilde p \gamma} \ln \left(\frac{\tilde p\gamma|S|}{2d}\right) +
\frac{4 d}{\tilde p \gamma}\ln(n) =
\frac{4 d}{\tilde p \gamma}\ln \left(\frac{\tilde p \gamma|S| n}{2d}\right) \implies
(\tilde p \gamma)^2 > \frac{4 \tilde p \gamma d}{n}  \ln\left(\frac{\tilde p\gamma |S| n}{2d}\right)$.
Taking the square root and substituting for $\epsilon$ on the right hand side yields
$\tilde p\gamma > 2 \epsilon$ and the claim follows.

Part (2) follows directly from part (1)

Part (3): $n>n_0 \ge  \frac{4e^2d}{\tilde p \gamma |S|} \ge
\frac{4e^2d}{\tilde p \gamma |S|^2}  \implies \sqrt{n} >
\frac{2ed}{\sqrt{\tilde p \gamma d} |S|}$. In addition $n>
\frac{4e^2d}{\tilde p \gamma |S|} > \frac{2de}{\tilde p\gamma|S|}$
which implies $1 < \ln\left(\frac{\tilde p\gamma |S| n}{2d}\right)$.
Combining these two inequalities we get: $\sqrt{n} >
\frac{2ed}{\sqrt{\tilde p \gamma d} \sqrt{\ln\left(\frac{\tilde
p\gamma |S| n}{2d}\right)}|S|}$. Multiplying both sides by $\sqrt{n}
\ $ implies $\ \ n > \frac{2ed\sqrt{n}}{\sqrt{\tilde p \gamma d}
\sqrt{\ln\left(\frac{\tilde p\gamma |S| n}{2d}\right)}|S|} =
\frac{2ed}{\epsilon|S|}.$

{\bf QED}

Using these parameters we set $\hat M(t) = \bar M_q^\epsilon(t)$. Our main  result is:

\bthm\label{main thm}{\bf (Main Theorem)}
The mechanism $\hat M(t)$ is ex-post Nash truthful and, in addition, it $ \ 6 \sqrt{\frac{d}{\tilde p \gamma n}} \sqrt{\ln \left(\frac{n\tilde p \gamma|S|}{2d}\right)}$-implements $F$ in ex-post Nash equilibrium, for $n > n_0$.
If agents have private reactions the mechanism is strictly truthful and $ \ 6 \sqrt{\frac{d}{\tilde p \gamma n}} \sqrt{\ln \left(\frac{n\tilde p \gamma|S|}{2d}\right)}$-implements $F$ in strictly dominant strategies.
\ethm

Recall that for ex-post Nash implementation we only need to show that one ex-post Nash equilibrium yields the desired outcome.

{\bf Proof}: Given the choice of parameters $\epsilon$ and $q$ then, Theorem \ref{theorem1} guarantees that $\hat M(t)$ is ex-post Nash truthful (and truthful whenever reactions are private). Therefore, it is sufficient to show that for any type vector $t$,
$$\ E_{\hat M(t)}(F(t,s)) \ge \max_s F(t,s) -  6 \sqrt{\frac{d}{\tilde p \gamma n}} \sqrt{\ln \left(\frac{n\tilde p \gamma|S|}{2d}\right)}.$$

Note that as $F$ is positive,  $E_{M^P(t)}[F(t,s)] \geq 0$ and so
\begin{eqnarray*}
E_{\hat M(t)}[F(t,s)] & \geq & (1-q) E_{M^{\frac{\epsilon}{2d}}(t)}[F(t,s)].
\end{eqnarray*}
By part (3) of Lemma \ref{lemma5} we are guaranteed that the condition on the size of of the population of Lemma \ref{lemma3} holds and so we can apply Lemma \ref{lemma3} to conclude that:
$$
E_{\hat M(t)}[F(t,s)] \geq (1-q) \left(\max_s F(t,s) - \frac{4d}{n\epsilon}\ln\left(\frac{n\epsilon|S|}{2d}\right)\right).$$

We substitute $q$ with $\frac{2\epsilon}{\tilde p \gamma}$ and recall that $\max_s F(t,s) \le 1$. In addition, part (1) of Lemma \ref{lemma5} asserts that  $\frac{2\epsilon}{\tilde p \gamma} <1$. Therefore
\begin{eqnarray*}
E_{\hat M(t)}[F(t,s)] \geq \max_s F(t,s) - \frac{2\epsilon}{\tilde p \gamma} - \frac{4d}{n\epsilon}\ln\left(\frac{n\epsilon|S|}{2d}\right) \ge
\max_s F(t,s) - \frac{2\epsilon}{\tilde p \gamma} -
\frac{4d}{n\epsilon}\ln\left(\frac{n\tilde p \gamma|S|}{2d}\right),
\end{eqnarray*}
where the last inequality is based on the fact $\epsilon < \tilde p \gamma$, which is guaranteed by part (2) of Lemma \ref{lemma5}.
Substituting $\epsilon$ for
$\sqrt{\frac{\tilde p \gamma d }{n}}\sqrt{\ln \left(\frac{n\tilde p \gamma|S|}{2d}\right)}$
we conclude that
$$ E_{\hat M(t)}[F(t,s)]  \geq \max_s F(t,s) - 2\sqrt{\frac{d}{\tilde p \gamma n}} \sqrt{\ln \left(\frac{n\tilde p \gamma|S|}{2d}\right)} - 4 \sqrt{\frac{d}{\tilde p \gamma n}} \sqrt{\ln \left(\frac{n\tilde p \gamma|S|}{2d}\right)}$$
and the result follows.

{\bf QED}

One particular case of interest is the commitment mechanism $M^U$, where $U$ is the uniform distribution over the set $S$:

\bcor\label{cor2}
Let $n_0$ be the minimal integer satisfying
$n_0 \ge \max\{ \frac{8\tilde d|S|}{\gamma} \ln \left(\frac{\gamma}{2d}\right),
\frac{4e^2d}{\gamma |S|} \} \ $
and $\ \frac{n_0}{\ln(n_0)} > \frac{8 d|S|}{\gamma}$.
Then the mechanism $\hat M^U(t)$ $\ \ \ 6 \sqrt{\frac{d|S|}{\gamma n}} \sqrt{\ln \left(\frac{n\gamma}{2d}\right)}$-implements $F$, in ex-post Nash equilibrium, for all $n > n_0$.
If agents have private reactions the mechanism $\hat M^U(t)$ $\ \ \ 6 \sqrt{\frac{d|S|}{\gamma n}} \sqrt{\ln \left(\frac{n\gamma}{2d}\right)}$-implements $F$ in strictly dominant strategies.
\ecor

{\bf Proof}:
 $P=U$ implies that the minimal probability is $\tilde p = \frac{1}{|S|}$. Plugging this into Theorem \ref{main thm} gives the result.

{\bf QED}

Holding the parameters of the environment $d,\gamma, |S|$ fixed the
approximation
inaccuracy of our mechanism converges to zero at a rate
of $\sqrt{\frac{\ln (n)}{n}}$.

In summary, by concatenating the exponential mechanism, where
truthfulness is
$\epsilon$-dominant with the commitment mechanism we
obtain a strictly truthful mechanism. In fact, this would hold true
for any mechanism where truthfulness is  $\epsilon$-dominant not
only the exponential mechanism.

\section{Applications}

We now turn to demonstrate the generic results in two concrete applications.

\subsection{Monopolist Pricing}

A monopolist producing digital goods, for which the marginal cost of
production is zero,
faces a set of indistinguishable buyers. Each
buyer has a unit demand with a valuation in the unit interval. Agents are arranged in (mutually exclusive) cohorts and the valuations of cohort
members are correlated. Each agent receives a private signal and her valuation is uniquely determined by the signals of all her cohort members. The monopolist wants to set a uniform price in order to maximize her average revenue per user.%
\footnote{To make this more concrete one can think of the challenge of pricing a fire insurance policy to apartment owners. Each apartment building is a cohort that shares the same risk and once the risk is determined (via aggregation of agents' signals) each agent has a private valuation for the insurance.}

Assume there are $N\cdot D$ agents, with agents labeled $(n,d)$ (the
$d^{th}$ agent in the $n^{th}$ cohort). Agent
$(n,d)$ receives a signal $X_d^n \in \real$ and we denote a cohort's vector of signals by $X^n=\{X^n_d\}_{d=1}^D$. We assume that the valuation of an agent, $V^n_d$, is uniquely determined by the signals of her cohort members; $V^n_d = V^n_d(X^n)$.

We assume that each agent's signal is informative in the sense that $V^n_d(X^n) > V^n_d(\hat
X^n)$ whenever $X^n > \hat X^n$ (in each coordinate a weak
inequality holds and for at least one of the coordinates a strong
inequality holds). That is, whenever an individual's signal
increases the valuation of each of her cohort members increases.

Let $R_{(n,d)}=\{\mbox{`Buy'}, \mbox{`Not buy'}\}$ be the set of reactions for agent $(n,d)$.

The utility of $(n,d)$, given the vector of signals $X= \{X^n\}_{n=1}^N = \{\{X^n_d\}_{d=1}^D\}_{n=1}^N$, and the price $p$, is
$$u_{(n,d)}(X,p,r_{(n,d)}) = \left\{\begin{array}{cl} V_d^n(X^n) -p & \mbox{if $r_{(n,d)} = \mbox{`Buy'}$,} \\ 0 & \mbox{if $r_{(n,d)} = \mbox{`Not buy'}$.} \end{array}\right.$$

We assume that all valuations are restricted to the unit interval, prices are restricted to some finite grid $S =S_m =  \{0,\frac{1}{m},\frac{2}{m},\ldots,1\}$ (hence, $|S|=m+1$), and
$X_d^n$ takes on only finitely many values. We assume the price grid is fine enough so that for any two vectors $X^n > \hat X^n$ there exists some price $p \in S$ such that $E(V^n|X^n) > p+\frac{2}{m}>p > E(V^n|\hat X^n)$. Therefore for vector of announcements there exists a maximal price for which optimal reaction is Buy. For that price, if an agent announces a lower value then the best reaction would be Not Buy, which will yield a loss of $\frac{1}{m}$ at least. Similarly, there exists the lowest price for which the optimal reaction is Not Buy. Announcing a higher value will result in the optimal reaction being Buy, which yields a loss of $\frac{1}{m}$ at least. We conclude that the gap is  $\gamma = \frac{1}{m}$.

The monopolist wants to maximize $F(t,p) = \frac{p}{ND}\cdot|\{(n,d): V^n_d(X^n) > p \}|$, the average revenue per buyer. Note that a unilateral change in the type of one agent may change at most the buying behavior of the $D$ members in her cohort, resulting in a change of at most $\frac{pD}{ND} \leq \frac{D}{ND}$ in the average revenue. As the population size is $ND$ we conclude that $F$ is $D$-sensitive.

Let $M_{dg}$ be a mechanism as in Corollary \ref{cor2}, where a Uniform Commitment mechanism is used:

 \bcor
For any $D$ there exists some $N_0$ such that for all $N>N_0$ the mechanism $M_{dg}$ $\ O(\sqrt{\frac{m^2}{N} \ln(\frac{N}{m})})$-implements $F$ in ex-post Nash equilibrium.
 \ecor

\remove{
This technique can extend to valuations and prices that are not
restricted to
the grid but can take any value in the unit interval.
For this setting let $M'_{dg}$ be the mechanism $M_{dg}$ applied to
the rounded announcements. That is we replace each announcement
$b_i$ with the the largest value in the grid $S_m$ not exceeding it
and apply $M_{dg}$. The loss of $M'_{dg}$ is that of $M_{dg}$ plus
the effect of discretization, which adds to $F$ at most
$\frac{1}{m}$, and hence we get that $M'_{dg}$ is an
$O(\sqrt{\frac{m^2}{N} \ln(N/m)} + \frac{1}{m})$-implementation.
Setting $m=(N/\ln N)^{1/4}$ we get that $M'_{dg}$ is an
$O\left((\ln(N) /N)^{1/4}\right)$-implementation for $F$ in ex-post
Nash equilibrium.
}

The literature on optimal pricing in this setting has so far
concentrated on the private values case and has provided
better approximations. For example, Balcan et al.~\cite{Balcan05},
using sampling techniques from Machine Learning, provide a mechanism
that $O(\frac{1}{\sqrt{n}})$-implements the maximal revenue without any restrictions to a grid.

\paragraph{A Multi Parameter Extension:} In the above setting we assumed a simple single-parameter type space. However, the
technique provided does not hinge on this. In particular, it extends to more complex settings where agents have a
multi-parameter type space. More concretely, consider a monopolist
that produces $G$ types of digital goods, each with zero marginal
cost for production. There are $N$ buyers, where each buyer assigns
a value, in some bounded interval, to each subset of the $G$ goods (agents want at most a singe unit of each good).
The monopolist sets $G$ prices, one for each good, and once prices are set each agent chooses his optimal bundle. The challenge of the monopolist is to
maximize the average revenue per buyer. In this model types are
sufficiently diverse. In fact, for any two types there exists a
price vector that yields different optimal consumptions.
Therefore, the scheme we provide applies just as well to this
setting.

\subsection{Facility Location}
\label{sec:facility location}

Consider a population of $n$ agents located on the unit interval. An
agent's location is private information and a social planner needs
to locate
$K$ similar facilities in order to minimize the average distance agents travel to the nearest facility.%
\footnote{For expositional reasons we restricting attention to the
unit interval and to the average travel distance. Similar results
can be obtained for other sets in $\real^2$ and other metrics, such
as distance squared.} We assume each agent wants to minimize her
distance to the facility that services her. In particular, this
entails that values (and reactions) are private. We furthermore assume that agent and facility locations are all restricted
to a fixed finite grid on the unit interval, $L= L(m) =
\{0,\frac{1}{m},\frac{2}{m},\ldots,1\}$. Using the notation of
previous sections, let $T_i=L$, $S = L^K$, and let $R_i=L$. The
utility of agent $i$ is
$$u_i(t_i,s,r_i) = \left\{\begin{array}{cl} -|t_i-r_i| & \mbox{if $r_i\in s$,} \\
-1 & \mbox{otherwise.}\end{array}\right. $$ Hence, $r_i(b_i,s)$ is
the facility closest to the locations of the facility in $s$ closest
to $b_i$. Let $F(t,s) = \frac{1}{n} \sum_{i=1}^n
u_i(t_i,s,r_i(t_i,s))$ be the social utility function, which is
$1$-sensitive (i.e., $d=1$).

First, consider the uniform commitment mechanism $\hat M^U$, which
is based on the uniform distribution over $S$ for the commitment
mechanism. Now consider the mechanism $\hat M_{LOC1}$, based on the
uniform commitment mechanism, as in Corollary \ref{cor2}

\bcor $\exists n_0$ such that $\forall n>n_0$ the mechanism
$\hat M_{LOC1}$  $\ 6 \sqrt{\frac{m(m+1)^K}{n}} \sqrt{\ln
\left(\frac{n}{2m}\right)}$- implements the optimal location in
strictly dominant strategies.
\ecor

{\bf Proof}: Note that $\gamma =\frac{1}{m}$, $|S|=(m+1)^K$ and the
proof follows immediately from  Theorem \ref{main thm}.

{\bf QED}

Now consider an alternative commitment mechanism. Consider the
distribution  $P$, over $S=L^K$, which chooses uniformly among all
the following alternatives - placing one facility in location
$\frac{j}{m}$ and the remaining $K-1$ facilities in location
$\frac{j+1}{m}$, where $j=0,\ldots,m-1$. Note that for any $i$, any
pair $b_i \not = t_i$ is separated by at least one alternative in
this set. For this mechanism $\tilde p = \frac{1}{m}$. Now consider
the mechanism $\hat M_{LOC2}$, based on the commitment mechanism,
$M^P$:

\bcor $\exists n_0$ such that $\forall n>n_0$ $\hat M_{LOC2}$ $ \  \frac{6m}{\sqrt{n}} \sqrt{\ln
\left(\frac{n(m+1)^K}{2m^2}\right)}$-implements the optimal
location in
strictly dominant strategies.
 \ecor

{\bf Proof:} In analogy to the proof of Theorem~\ref{main thm},
setting $\epsilon = \frac{1}{m\sqrt n}\sqrt{\ln
\left(\frac{n(m+1)^K}{2m^2}\right)}$ and $q=2\epsilon m^2$.

{\bf QED}

For both mechanisms the approximation error converges to zero at a
rate proportional to $1/\sqrt{n}$ as society grows. In addition, the approximation error of both mechanisms grows as the grid size, $m$, grows. However in the second mechanism approximation deteriorates at a substantially slower rate.

\section{Large Type Sets}

The arguments underlying the generic approximate optimal mechanism for a finite number of social alternatives do not generally extend to models where the type set is large. However, in concrete models, where additional structure is assumed, such an extension may be possible. We demonstrate this in the facility location problem introduced in the previous section.

As before, we assume that each player is located on the unit interval and that her location is private information. Formally, set $T=[0,1]$. A mechanism must (randomly) decide on the location of $K$ facilities in the unit interval. Let $S = [0,1]^K$ and consider the standard Borel $\sigma$-algebra which we denote $\cal S$. The objective of the designer  is to minimize the average distance a player must travel to a facility. Formally, the designer seeks to minimize $F(t,s)=\frac{1}{n}\sum_{i=1}^{n} |t_i- r_i(t_i,s) |$, where $r_i(t_i,s)$ denotes the facility in $s$ that is closest to $t_i$.

We use a continuous version of the {\em Exponential Mechanism}, where the probability of any event $\hat S \in \cal S$ is given by:
 $$M^{\epsilon}(t)(\hat S) = \frac{\int_{\hat S} e^{n\epsilon F(t,s)} \mathrm{d}s}{\int_{S} e^{n\epsilon F(t,s)} \mathrm{d}s} \ \ \forall \hat S \in \cal S.$$
We say that a mechanism $M$ provides $\epsilon$-differential privacy if $\frac{M(t)(\hat S)}{M(\hat t)(\hat S)} \leq e^\epsilon \ \  \forall \hat S \in \cal S$ and for any pair of type tuples, $t$ and $\hat t$, that differ on a single entry. McSherry and Talwar \cite{MT07} prove the following (which is analogous to lemma ~\ref{lemma1}):

\blem[McSherry and Talwar \cite{MT07}]\label{lemma1continuous}
If $F$ is $d$-sensitive then $M^{\frac{\epsilon}{2d}}(t)$ preserves $\epsilon$-differential privacy.
\elem

The proof is identical to that of Lemma \ref{lemma1} and is
therefore omitted.

\subsection{The approximation accuracy of the Exponential Mechanism}

The solution concept we pursue in this section is deletion of
dominated strategies. In fact, what we show in the sequel is that being truthful dominates significantly mis-reporting one's type. Thus, deletion of dominated strategies implies that agents resort to
strategies that are `almost' truthful.

Consequently, we turn to study the approximation accuracy of the
Exponential Mechanism whenever agents slightly mis-report their
types. We begin by considering truthful agents.

To state the next lemma we introduce the following notation. For
$0\leq \alpha \leq 1$ let $S_\alpha = S_\alpha(t) = \{\bar s\in S:
F(t,\bar s) \geq \max_s F(t,s) -\alpha\}$, and $\bar S_\alpha = \bar
S_\alpha(t) = S \setminus S_\alpha$. Let $\mu$ denote the uniform
probability over $\cal S$.

\blem[McSherry and Talwar \cite{MT07}]\label{lemma3continuous}
If $\alpha \geq \frac{2d}{n\epsilon}\ln\left(\frac{\max_s
F(t,s)}{\alpha \mu(S_\alpha)}\right)$ then $E_{M^{\frac{\epsilon}{2d}}(t)}[F(t,s)] \geq$ $\max_s F(t,s) -
3\alpha$. \elem

We include the proof for completeness.

{\bf Proof}: Note first that
\begin{eqnarray*}
M^{\frac{\epsilon}{2d}}(t)(\bar S_{2\alpha}) & \leq & \frac{M^{\frac{\epsilon}{2d}}(t)(\bar S_{2\alpha}) }{M^{\frac{\epsilon}{2d}}(t)(S_\alpha) }
= \frac{\int_{\bar S_{2\alpha}} e^\frac{n\epsilon F(t,\bar s)}{2d} \mathrm{d} \bar s}{\int_{S_\alpha} e^\frac{n\epsilon F(t,\bar s)}{2d} \mathrm{d} \bar s}
\\
&\leq & \frac{\int_{\bar S_{2\alpha}} e^\frac{n\epsilon (\max_s F(t,s) -2\alpha)}{2d} \mathrm{d} \bar s}{\int_{S_\alpha} e^\frac{n\epsilon (\max_s F(t,s) -\alpha)}{2d} \mathrm{d} \bar s}
= e^\frac{-n\epsilon\alpha}{2d}\cdot \frac{\mu(\bar S_{2\alpha})}{\mu(S_\alpha)}   \leq  \frac{e^\frac{-n\epsilon\alpha}{2d}}{\mu(S_\alpha)},
\end{eqnarray*}
where the first inequality follows from $M^{\frac{\epsilon}{2d}}(t)(S_\alpha) \leq 1$, the second inequality follows from the definition of $\bar S_{2\alpha}$ and $S_\alpha$, and the third inequality follows from $\mu(\bar S_{2\alpha}) \leq 1$. Hence, we get that $M^{\frac{\epsilon}{2d}}(t)$ returns $s\in S_{2\alpha}$ with probability at least $1- \frac{e^\frac{-n\epsilon\alpha}{2d}}{\mu(S_\alpha)} \geq 1- \frac{\alpha}{\max_s F(t,s)}$. Hence,
$$E_{M^{\frac{\epsilon}{2d}}(t)}[F(t,s)] \geq (\max_s F(t,s) - 2\alpha)(1- \frac{\alpha}{\max_s F(t,s)}) \geq \max_s F(t,s) - 3\alpha.$$

{\bf QED}

This result enables us to prove the following:

\bcor\label{corollaryfacility}
$E_{M^{\frac{\epsilon}{2}}(t)}[F(t,s)] \geq \max_s F(t,s) - \frac{6}{n\epsilon}\ln \left(e+(n\epsilon)^{K+1}\right)$.
\ecor

{\bf Proof:} Fix a tuple of players' locations $t \in T^n$ and let $s$ denote the alternative in $S$ that minimizes $F(t,s)$.  For any $\alpha> 0$, if $\hat s \in [0,1]^K$ satisfies $\max_k|\hat s_k -s_k|<\alpha$ then $\hat s \in  S_\alpha$. To see this note that
\begin{eqnarray*}
F(t,s') & = & \frac{1}{n}\sum_{i=1}^{n}u_i(t_i,s',r_i(t_i,s')) \\
& \leq & \frac{1}{n}\sum_{i=1}^{n}u_i(t_i,s',r_i(t_i,s)) \\
& \leq & \frac{1}{n}\sum_{i=1}^{n}\left(u_i(t_i,s,r_i(t_i,s)) + \alpha\right) \\
& \leq & F(t,s') + \alpha.
\end{eqnarray*}

Therefore, whenever  $\alpha \leq 0.5$, $\ \mu(S_\alpha) \geq \alpha^K$.

Set $\alpha = \frac{2}{n\epsilon}\ln \left(e+(n\epsilon)^{K+1}\right)$. We argue that
$\alpha \geq \frac{2}{n\epsilon}\ln\left(\frac{\max_s F(t,s)}{\alpha \mu(S_\alpha)}\right)$ which implies that we can apply Lemma~\ref{lemma3continuous}. To see this recall that  $\max_s F(t,s) \leq 1$, and using our bound on $\mu(S_\alpha)$ it suffices to show that $(n\epsilon)^{K+1}  \geq 1/\alpha^{K+1}$, which indeed is the case as $\alpha \geq 1/n\epsilon$. By Lemma~\ref{lemma3continuous}  $E_{M^{\frac{\epsilon}{2}}(t)}[F(t,s)] \geq \max_s F(t,s) - \frac{6}{n\epsilon}\ln \left(e+(n\epsilon)^{K+1}\right)$, as required.

{\bf QED}

We now turn to analyze the case where agents misreport their
location.

\blem\label{lemma10} \
\begin{itemize}
\item
$|F(b_i,t_{-i},s) - F(t_i,t_{-i},s)| \leq \frac{1}{n}|t_i-b_i|$; and
\item
$|F(b,s) - F(t,s)| \leq max_i |t_i-b_i|$.
\end{itemize}
\elem

{\bf Proof:}
To derive the first part note that
\begin{eqnarray*}
-u_i(b_i,s,r_i(b_i,s)) & = & |b_i-r_i(b_i,s)| \leq |b_i-r_i(t_i,s)| = |b_i-t_i+t_i-r_i(t_i,s)| \\
& \leq & |b_i-t_i|+|t_i-r_i(t_i,s)| = |b_i-t_i| - u_i(t_i,s,r_i(t_i,s)),
\end{eqnarray*}
and (by a similar analysis) $-u_i(t_i,s,r_i(t_i,s))  \leq |b_i-t_i| - u_i(b_i,s,r_i(b_i,s))$. Hence, $|u_i(b_i,s,r_i(b_i,s)) - u_i(t_i,s,r_i(t_i,s))| \leq |b_i-t_i|$ and we get that
$$ |F(b_i,t_{-i},s) - F(t_i,t_{-i},s)| =
\frac{1}{n} |u_i(b_i,s,r_i(b_i,s)) - u_i(t_i,s,r_i(t_i,s))| \leq \frac{1}{n}|t_i-b_i|.$$

The second part follows by iteratively applying the first part for $n$
times.

{\bf QED}

\blem\label{lemma11} If $\ |b_i-t_i| \leq \beta$ for all $i$ then
$|\max_s F(t,s) - \max_s F(b,s)| \le \beta$.
\elem

{\bf Proof:} Let $s_t \in \argmax_s\{F(t,s)\}$ and $s_b \in
\argmax_s\{F(b,s)\}$. Using triangle inequality, $|t_i-b_i| + |b_i-r_i(b_i,s_b)| \geq |t_i-r_i(b_i,s_b)|$, and noting that $ |t_i-r_i(b_i,s_b)| \geq |t_i-r_i(t_i,s_b)|$ we get that $$|t_i-b_i| + |b_i-r_i(b_i,s_b)| \geq |t_i-r_i(t_i,s_b)|.$$ Hence,
$$
{1 \over n}\sum_{i=1}^n -\left(|t_i-b_i| + |b_i-r_i(b_i,s_b)|\right) \leq
{1 \over n}\sum_{i=1}^n -|t_i-r_i(t_i,s_b)| = F(t,s_b) \leq F(t,s_t).
$$
Noting that ${1 \over n}\sum_{i=1}^n -|b_i-r_i(b_i,s_b)| = F(b,s_b)$ we get that
\begin{equation}\label{eq:large1}
F(b,s_b)-F(t,s_t) \leq {1 \over n}\sum_{i=1}^n |t_i-b_i| \leq \beta.
\end{equation}
A similar argument yields
\begin{equation}\label{eq:large2}
F(t,s_t)-F(b,s_b)  \leq \beta.
\end{equation}
Combining inequalities \ref{eq:large1} and \ref{eq:large2} we conclude that
$$|\max_s F(t,s) - \max_s F(b,s)|  = |F(t,s_t)-F(b,s_b)| \leq\beta,$$ as claimed.

{\bf QED}

\blem\label{lemma12}
If $\ |b_i-t_i| \leq \beta$ for all $i$ then
$E_{M^{\frac{\epsilon}{2}}(b)}[F(t,s)] \geq \max_s F(t,s) - 2\beta -
\frac{6}{n\epsilon}\ln \left(e+(n\epsilon)^{K+1}\right)$ \elem

{\bf Proof:} For any finite set of locations $s \subset [0,1]$, traveling from $t_i$ to the point closest to $t_i$ in $s$ is not longer than a  taking a detour via $b_i$, and then traveling from $b_i$ to the point closest to $b_i$ in $s$, i.e., $|t_i-r_i(t_i,s)| \leq |t_i-b_i| + |b_i-r_i(b_i,s)|$.
Therefore, $$E_{M^{\frac{\epsilon}{2}}(b)}[F(t,s)] \ge
E_{M^{\frac{\epsilon}{2}}(b)}[F(b,s)] -{1 \over n}\sum_{i=1}^n
|t_i-b_i| \ge E_{M^{\frac{\epsilon}{2}}(b)}[F(b,s)] - \beta.$$
By Corollary~\ref{corollaryfacility},
$$E_{M^{\frac{\epsilon}{2}}(b)}[F(b,s)] \geq \max_s F(b,s) -
\frac{6}{n\epsilon}\ln \left(e+(n\epsilon)^{K+1}\right).$$
By Lemma~\ref{lemma11}, $$\max_s F(b,s) \ge \max_s F(t,s) - \beta.$$

Combining all three inequalities above gives:
$$E_{M^{\frac{\epsilon}{2}}(b)}[F(t,s)] \ge \max_s F(t,s) - 2\beta -
\frac{6}{n\epsilon}\ln \left(e+(n\epsilon)^{K+1}\right),$$ as
claimed.

{\bf QED}

\subsection{Deviations from truthfulness in the Exponential
Mechanism}

We bound the potential gain of an agent located at $t_i$ who
reports $b_i$:

\blem\label{lemma6cont} Using the Exponential Mechanism for the facility location problem, if
$\epsilon \leq 1$ then for any $i$, any $b_i,t_i \in T_i$ and any
$t_{-i} \in T_{-i}$,$$E_{M^{\frac{\epsilon}{2}}(b_i,t_{-i})}
[u_i(t_i,s,r_i(t_i,s))] -
E_{M^{\frac{\epsilon}{2}}(t_i,t_{-i})}[u_i(t_i,s,r_i(t_i,s))] \le
2\epsilon|t_i-b_i|.$$ \elem

{\bf Proof}:
By Lemma~\ref{lemma10}, $|F(b_i,t_{-i},s) - F(t_i,t_{-i},s)| \leq \frac{1}{n}|t_i-b_i|$.
Plugging this into the definition of the Exponential Mechanism we
get:
\begin{eqnarray*}
E_{M^{\frac{\epsilon}{2}}(b_i,t_{-i})}[u_i(t_i,s,r_i(t_i,s)]
& = & \int_{s\in S} u_i(t_i,s,r_i(t_i,s)) \ \mathrm{d} M^{\frac{\epsilon}{2}}(b_i,t_{-i})(s) \\
& = & \int_{s\in S} u_i(t_i,s,r_i(t_i,s)) \frac{e^{\frac{n\epsilon}{2} F(b_i,t_{-i},s)}}{\int_{s'\in S}{e^{\frac{n\epsilon}{2} F(b_i,t_{-i},s')}}\ \mathrm{d}s'}\ \mathrm{d}s \\
& \leq & \int_{s\in S} u_i(t_i,s,r_i(t_i,s)) \frac{e^{\frac{n\epsilon}{2} \left(F(t_i,t_{-i},s) + \frac{|t_i-b_i|}{n}\right)}}{\int_{s'\in S}{e^{\frac{n\epsilon}{2} \left(F(t_i,t_{-i},s') -\frac{|t_i-b_i|}{n}\right)}}\ \ \mathrm{d} s'} \ \mathrm{d}s\\
& = & e^{\epsilon|t_i-b_i|} \int_{s\in S} u_i(t_i,s,r_i(t_i,s)) \ \mathrm{d} \ M^{\frac{\epsilon}{2}}(t_i,t_{-i})(s) \\
& = & e^{\epsilon|t_i-b_i|} E_{M^{\frac{\epsilon}{2}}(t_i,t_{-i})}
[u_i(t_i,s,r_i(t_i,s)],
\end{eqnarray*}

The proof is completed by noting that as $|t_i-b_i| \leq 1$ and
$\epsilon < 1$, $\ e^{\epsilon|t_i-b_i|} \leq 1+2
\epsilon|t_i-b_i|$.

{\bf QED}

\subsection{A commitment mechanism}
\label{sec:commitment large type sets}

Consider a commitment mechanism induced by the following
distribution $P$ over the set $S=[0,1]^K$: First, choose a uniformly
a random integer $X \in \{1,2,3,\ldots,{\bar m}\}$, where the
parameter $\bar m$ will be set below. Next choose a number $Y$,
randomly and uniformly, from the interval $[0,2^X-1]$. Now let $s$
be the alternative where one facility is located at $\frac{Y}{2^X}$
and the other $K-1$ facilities at $\frac{Y+1}{2^X}$.

\blem If $|b_i-t_i| \geq 2^{-(\bar m-1)}$ then $\ E_{M^P(t_i,t_{-i})}[u_i(t_i,s)] \ge  E_{M^P(b_i,t_{-i})}[u_i(t_i,s)] + \frac{|t_i-b_i|^2}{8\bar m}$. \elem

{\bf Proof:} We first consider the case $b_i\leq t_i-2^{-(\bar m-1)}$. Assume $X$ and $Y$ are chosen such that $\frac{1}{2^X} < \frac{|t_i-b_i|}{2} \le \frac{2}{2^X}$ and $\frac{Y}{2^X} \in [b_i, \frac{b_i+t_i}{2}]$. As a result the facility assigned to $i$, whenever she announces $b_i$ is located at $\frac{Y}{2^X}$. However, if she announces her true location,  $t_i$, she is assigned a facility located at $\frac{Y+1}{2^X}$. Consequently, $u_i(t_i,s,r_i(t_i,s)) \ge u_i(t_i,s,r_i(b_i,s)) + \frac{1}{2^X} \ge u_i(t_i,s,r_i(b_i,s)) +
\frac{(t_i-b_i)}{4}$. In words, for the specific choice of $X$ and
$Y$ misreporting one's type leads to a loss exceeding
$\frac{(t_i-b_i)}{4}$.

The probability of choosing the unique $X$ satisfying  $\frac{1}{2^X} <
\frac{|t_i-b_i|}{2} \le \frac{2}{2^X}$ is $1/{\bar m}$. Conditional on this event,
the probability of choosing $Y$ satisfying
$\frac{Y}{2^X} \in [b_i, \frac{b_i+t_i}{2}]$ is  $\frac{(t_i-b_i)}{2}$.
Since the mechanism is imposing, then for an arbitrary choice of $X$ and $Y$
misreporting is not profitable. Therefore, the expected loss
from misreporting exceeds $\frac{|t_i-b_i|^2}{8\bar m}$.

The proof of the complementary case, $b_i\geq t_i +2^{-(\bar m-1)}$,
uses similar arguments and is omitted.

{\bf QED}

\subsection{Implementation in undominated strategies}

As in the generic construction, let $\bar M_q^\epsilon(t) =
(1-q)M^{\frac{\epsilon}{2}}(t) + q M^P(t)$. Note that whenever $q
\frac{|t_i-b_i|^2}{8{\bar m}} \ge 2\epsilon|t_i-b_i|$ being truthful dominates any announcement satisfying $|b_i-t_i| \geq 2^{-(\bar m-1)}$. In particular, this holds whenever
$q \ge 16\epsilon\bar m 2^{\bar m}$.

Set $\epsilon = \frac{1}{n^{2/3}}\sqrt{K+1}$,
$\bar m = \lceil \log \left(\frac{n^{1/3}}{6\sqrt{K+1}\ln n}\right) \rceil$, and
$q=16\epsilon {\bar m} 2^{\bar m}$ and denote by $\hat M_{LOC3} = \bar M_q^\epsilon$ for this choice of parameters.

\bthm There exists $n_0=n_0(K)$ such that $\hat M_{LOC3}$ $\
\frac{32\sqrt{K+1}}{n^{1/3}}\ln n$-implements $F$ in undominated
strategies for all $n>n_0$. \ethm

{\bf Proof:} We first observe that as there exists $n_q=n_q(K)$ such that $q < 1$ for
all $n > n_q$ and hence the mechanism is well defined.
This also implies that for agent $i$ reporting $b_i$ such that
$|b_i-t_i| \geq 2^{-(\bar m-1)}$ is dominated by reporting $t_i$.
Similarly, there exists $n_\alpha = n_\alpha(K)$ such that
$\alpha = \frac{2}{n\epsilon}\ln \left(e+(n\epsilon)^{K+1}\right) \leq 0.5$
for all $n > n_\alpha$ as required in the proof of Corollary~\ref{corollaryfacility}.
Finally, there exists $n_{\bar m} = n_{\bar m}(K)$ such that $\bar m \leq \ln n$
for all $n > n_{\bar m}$. In the following we will assume that $n > \max(n_q,n_\alpha,n_{\bar m})$.

There are two sources for the additive error for $\hat M_{LOC3}$:
\begin{enumerate}
\item The commitment mechanism introduces an additive error of at most
$q = 8\epsilon {\bar m} 2^{\bar m}$. Noting that
$2^{\bar m} \leq 2\cdot \frac{n^{1/3}}{6\sqrt{K+1}\ln n}$
and substituting for $\epsilon$ we get that
$q \leq \frac{8}{3}\frac{\ln n}{n^{1/3}} \ \le \ \frac{2\sqrt{K+1}}{n^{1/3}}\ln n$.

\item The exponential mechanism introduces an additive error of ${2\over 2^{-(\bar m -1)}} +
\frac{6}{n\epsilon}\ln \left(e+(n\epsilon)^{K+1}\right)$ (see Lemma~\ref{lemma12}). Note that  $\frac{6}{n\epsilon}\ln
\left(e+(n\epsilon)^{K+1}\right) \leq \frac{6(K+1)}{n\epsilon}\ln
\left(e+n\epsilon\right)$ and substituting for $\epsilon$, we get
that there exists $n_1=n_1(K)$ such that for all $n > n_1$ this
additive error is bounded by ${2\over 2^{-(\bar m -1)}} +
\frac{6\sqrt{K+1}}{n^{1/3}}\ln n$. In addition $2^{\bar m -1} = {1
\over 2}\cdot 2^{\bar m} \ge {1 \over 2} \frac{n^{1/3}}{6\sqrt{K+1}\ln
n}$ which implies that the error is bounded by $ 4\cdot\frac{6\sqrt{K+1}
}{n^{1/3}}\ln n + \frac{6\sqrt{K+1}}{n^{1/3}}\ln n =
\frac{30\sqrt{K+1}}{n^{1/3}}\ln n$.
\end{enumerate}

Setting $n_0 = \max(n_q,n_\alpha,n_{\bar m}, n_1)$, we get that for all $n > n_0$
the total additive error is bounded by
$$ \frac{2 \sqrt{K+1}}{n^{1/3}}\ln n + \frac{30 \sqrt{K+1}}{n^{1/3}}\ln n =
\frac{32 \sqrt{K+1}}{n^{1/3}}\ln n.$$

{\bf QED}

\section{Discussion}

The mechanisms proposed in this paper are based on two pillars -- a differentially private mechanism on the one hand and an imposing mechanism on the other hand. In the following we discuss the importance of each of these pillars for the results obtained. In addition, we discuss some of the limitations of our results.

\subsection{Is differential privacy sufficient?}

McSherry and Talwar \cite{MT07} observed that differential privacy is sufficient to yield approximate implementation in $\epsilon$-dominant strategies. However, as we show below, differential privacy does not generally imply implementation with a stronger solution concept.

Our example is a pricing mechanism that utilizes the exponential mechanism and hence yields an $\epsilon$-dominant implementation that (assuming parties act truthfully) well approximates the optimal revenue. However, there are dominant strategies in the example that involve mis-representation and lead to a significantly inferior revenue.

\bexa\label{example1} Consider a monopolist producing an unlimited supply digital good who faces $n$ buyers, each having a unit demand at a valuation that is either $0.5+\mu$ or $1+\mu$ where $0 < \mu < 0.5$.
The monopolist cannot distinguish among buyers and is restricted to choosing a price in the set $\{0.5,1\}$.
Assume the monopolist is interested in maximizing the average revenue per buyer.%
\footnote{We consider the average revenue per buyer as the objective
function, instead of the total revenue, in order to comply with the
requirement that the value of the objective function is restricted
to the unit interval.}
The optimal outcome for the auctioneer is hence $$
OPT(\bar t) = \frac{\max_{s\in \{0.5,1\}} (s\cdot\left|\{i:t_i \geq
s\}\right|)}{n}.$$

If the monopolist uses the appropriate exponential mechanism then it is $\epsilon$-dominant for agents to
announce their valuation truthfully, resulting in an almost optimal revenue.
However, one should note that the probability that the exponential mechanism will choose the lower of the two prices increases with the number buyers that announce $0.5$. Hence, it is
{\bf dominant} for buyers to announce $0.5$. This may lead to
inferior results. In particular, whenever all agents value the good
at $1$ but announce $0.5$ the mechanism will choose the price $0.5$
with high probability, leading to an average revenue of $0.5$ per
buyer, which is half the optimal revenue per buyer.
\eexa

\subsection{Is imposition sufficient?}

It is tempting to think that our notion of imposition trivializes
the result, i.e., that, regardless of the usage of a
differentially-private mechanism, the ability to force agents to
react sub-optimally, according to their announced types, already
inflicts sufficient disutility that would deter untruthful
announcements. The next example demonstrates that such a naive
imposition is generally insufficient. Intuitively, the reason is
that for inducing both truthfulness and efficiency, one needs a
strong bound on an agent's benefit from mis-reporting: the utility
from mis-reporting should be smaller from the disutility from being
committed to a sub-optimal reaction.

\bexa\label{example3}
Consider a digital goods pricing problem with $n$ agents, where the valuation of each agent is either ${1 \over n}$ or $1+\mu$,
and the possible prices are ${1 \over n}$ and $1$. In this example the optimal price is $1$ whenever there exists an agent of type $1+\mu, \ \mu < 0.5$.

Consider the following mechanism: with high probability it implements the optimal price and with a low probability it uses an imposing mechanism. Note that the strategy to always announce a valuation of ${1 \over n}$ is a Nash equilibrium. This announcement is clearly optimal if an agent's valuation is indeed ${1 \over n}$. If an agent's valuation, on the other hand, is $1+\mu$, then complying with this strategy will result in a utility that is almost $1$, whereas deviating to truthful announcement will result in a price of $1$ with high probability, hence a utility of $\mu$.

Therefore, the monopolist's average revenue from a buyer is always ${1 \over n}$. This is substantially inferior to the optimal outcome, which could be as high as $1$,  whenever all agents are of the high type.
\eexa

The Nash equilibrium from example \ref{example3} survives even if we modify the mechanism to be fully imposing (i.e, it always imposes the optimal reaction). Thus, the above mentioned sub-optimality holds.

We believe that the notion of imposition is natural in many settings, and that to some extent imposition is already {\em implicitly} integrated into the mechanism design literature.
In fact, any mechanism that is not ex-post individually rational
imposes its outcome on the players: it imposes participation and ignores the possibility players have to `walk away' once the results are known. Moreover, models that involve transfers treat these as imposed reactions: once the social choice and transfers are determined, players must comply (consider taxation and auction payments as an example).


\subsection{Model Limitations}

There are three overarching limitations to the technique we present:
(1) The generic mechanism only works for objective functions which
are not sensitive, (2) We consider settings where the reaction set
of agents is rich enough, such that any pair of types can be
separated by the optimal reaction on at least one social
alternative; and (3) The size of the set of social alternatives
cannot grow too fast as the set of agents grows. We discuss these below.

\subsubsection{Low sensitivity of the objective function}

Many objective functions of interest are actually insensitive and comply with our requirements. Revenue in the setting of digital goods, and social welfare (i.e., sum of agents' valuations) are typical examples. We note that although we focused our attention on social functions whose sensitivity is constant (independent of $n$), one can apply Theorem~\ref{main thm} also in the case where $d=d(n)$ as long as $\frac{d(n)}{n} \rightarrow 0$ as $n\rightarrow \infty$.

There are, however, important settings where the objective function is sensitive and hence our techniques cannot be applied. An important example is that of revenue maximization in a single unit auction -- it is easy to come up with extreme settings where a change in a type of a single agent can drastically change the revenue outcome. Consider, e.g. the case where all agents value the good at zero, resulting in a maximal revenue of zero. A unilateral change in the valuation of any single agent from zero to one will change the maximal revenue from zero to one as well.

However, even in this case the domain of type profiles (valuation profiles) that demonstrate sensitivity is quite small -- for instance, if agents valuations are taken uniformly from $[0,1]$ then although the worst-case sensitivity of the maximal revenue is $1$, the 'typical' sensitivity would be of order $1/n$. In this case, the work of Nissim et al.~\cite{NRS07} may turn to be applicable, as it yields differentially private mechanisms where the deviation from the maximum depends on a local notion of sensitivity called {\em smooth sensitivity}. We leave the examination of this approach to future work.

\subsubsection{Rich reaction set}

The second limitation is the requirement that agents'
reaction sets are sufficiently rich. In fact, what we need for the
results to hold is that for any pair of types of an agent there
exists some social alternative for which the set of optimal reactions for
the first type is disjoint of the set of optimal reactions for the second
type. For example, in an auction setting, we require that for each
pair of agents' valuations the auctioneer can propose a price such
that one type will buy the good, while the other will refuse.

\subsubsection{Small number of social alternatives}

The approximation accuracy we achieve in Theorem~\ref{main thm} is proportional to $\sqrt{\frac{\ln(n\tilde p |S|)}{\tilde p n}}$. Note that $\tilde p > 1/|S|$. A naive use of the theorem yields accuracy $O(|S|\ln n/n)$, yielding meaningful approximation as long as $|S|$ (as a function of $n$) grows slower than $n/\ln n$.

As we have demonstrated in sections~\ref{sec:facility location}, one can sometimes design a commitment mechanism realizing a much bigger $\tilde p$, ideally independent of $|S|$. If that is the case, then Theorem~\ref{main thm} yields approximation error $O(\sqrt{\frac{\ln(n|S|)}{n}})$, allowing the number of social
alternatives to be as high as an exponential function of the number of agents.
For larger $S$,  the approximation error may not vanish as $n$ increases. Two interesting examples for such settings are matching problems, where each social alternative specifies the list of pairs, and multi unit auctions where the number of goods is half the number of bidders.


\subsection{Alternative mechanisms}

The framework we presented combines a differentially private
mechanism with
an imposing one. Our general results refer to a
`universal' construction of an imposing mechanism (the uniform one),
yet the specific examples we analyze demonstrate that imposing
mechanisms that are tailor made to the specific setting can improve
upon the results.

Similarly, it is not imperative to use the Exponential mechanism as
the first component, and other differentially-private mechanisms may
be adequate. In fact, the literature on differential privacy
provides various alternatives that may outperform the Exponential
mechanism, given a specific context. Some examples can be found in
Dwork et al.~\cite{DMNS06}, where the mechanism has a noisy
component that is calibrated to global sensitivity, or in Nissim et
al.~\cite{NRS07} where a similar noisy component is calibrated to
smooth sensitivity. The latter work also uses random sampling to
achieve similar properties. To learn more the reader is referred to
the recent survey of Dwork~\cite{Dwork09}.

\bibliographystyle{plain}

\end{document}